\documentclass[aps,pre,twocolumn]{revtex4}
\usepackage{graphicx}
\usepackage{amsmath}
\usepackage{braket}
\usepackage{hyperref}
\usepackage{subfigure}
\usepackage[usenames,dvipsnames]{color}

\newcommand{\ct}{\cite}

\newcommand{\be}{\begin{equation}}
\newcommand{\ee}{\end{equation}}
\newcommand{\ba}{\begin{eqnarray}}
\newcommand{\ea}{\end{eqnarray}}

\newcommand{\noi}{\noindent}

\begin{document}
\title{A simple experiment demonstrating the connection between chaos and music}
\author{Nitica Sakharwade}
\affiliation{Department of Physics, Indian Institute of Technology, 208016, Kanpur}
\author{Sayak Dasgupta}
\affiliation{Department of Physics, Indian Institute of Technology, 208016, Kanpur}
\begin{abstract}
In this paper we study the dripping pattern of a leaking faucet which we analyse on the basis of a musical procedure which we outline and match the power spectral density of these drops (which are recorded as noise signals over time) to verify the $\frac{1}{f^{\beta}}$ power law which characterizes the nature of such systems with non linear characteristics.
 \end{abstract}
\maketitle

\section{Introduction:}
\label{sec_Intro}
\noi
Human music contains patterns. For instance, harmony is formed by notes whose frequencies have a simple integer multiple relation. Other mathematical relations have been used to understand and model music, such as neural topography (Janata et al. 2002 \cite{Janata}) and orbifold space (Tymoczko 2006 \cite{tym}). In particular, the $1/f$ power law has been found useful to characterize different genres of music (Voss 1978, Hennig et al. 2011, Levitin et al. 2012 \cite{voss,Hennig,Levitin}).For a time-domain signal $V(t)$, the frequency-domain
power spectral density $S_{V}(\omega)$ is related to the time-domain auto-correlation function $h(t)$ by:
\begin{equation}
\label{eq.s_v}
S_{V}(\omega)=\int_{-\infty}^{\infty}h(t)e^{-i\omega t}dt
\end{equation}
where
\begin{equation}
\label{eq.h}
h(t)= \int_{-\infty}^{\infty}V^{*}(\tau)V(t+\tau)d\tau
\end{equation}
$S_{V}(\omega)$ is an indication of the correlation of $V(t)$, which can be obtained from its Fourier Transform. The $1/f$ power law describes a signal whose power spectral density $S_{V}(\omega)$, or $S_{V}(f)$), differ simply by a factor of $2\pi$, obeys the relation:
\begin{equation}
\label{eq.s_v_2}
S_{V}(f)=\frac{1}{f^{\beta}}
\end{equation}
For white noise, where V(t) has no temporal correlations,$\beta = 0$; for Brownian noise $\beta = 2$, which means that $V(t)$ is strongly correlated. We also see that in the case of the brown noise the correlation function $h(t)$ is proportional to $t$ which is expected as integrating the correlation function over time gives the rate of energy transmission or power which one can relate to the rate of diffusion in Brownian motion which varies as $t^{2}$. The extreme case of $\beta = 0$ which signifies no correlation is also logical as it is produced by an $h(t)$ which is a delta function mathematically signifying no correlation. Another important point to note is that the power law does not produce any quantity of dimension of time with which we can characterize the system. This is a property of fractal systems which suggests that no matter how small a time scale we choose to investigate the phenomenon over, the same power law results.\\ Voss \ct{voss}, Hennig \ct{Hennig}, and Levitin \ct{Levitin} have shown that in many musical pieces, from classical to rock music, the fluctuation of pitch (frequency), loudness and duration obeys $1/f$ power law. The exponent $\beta$ ranges from $0.4$ to $1.4$, depending on the composer and genre. The range of $\beta$ suggests that human music keeps a balance between predictability $(\beta = 2)$ and randomness $(\beta = 0)$. The $1/f$ relation is also observed in several natural phenomena, such as the frequency of earthquakes and the fluctuation of heart beat rate. The chaotic leaky faucet, used in a past UIUC Advanced Physics Lab course (\ct{course}) which we adapt in our experiment, may exhibit chaos in the time difference between successive drops produced because of the standing wave and damping in water (Martien \ct{martien}). The time difference follows the $1/ f$ power law, and is used as the $1/f$ noise source in this experiment.
\section{The experimental set up:}
\label{sec_setup}
\subsection{Apparatus}
\noi
1. Burette ($50 ml$)\\
2. Beaker ($100 ml$)\\
3. Buckets (which act as reservoirs)\\
4. Pipes\\
5. Screw clamps\\
6. Sealant\\
7. Metal bowl\\
8. Metal plate to act as a wave guide\\
9. Recording software (we use an audio recording software (garage band) which gives a high precision time resolution ($10^{-3}s$)).\\
\subsubsection*{Least Counts}
\noi
1. Timer on recording software = $10^{-3}$s\\
2. Burette = $0.1 ml$
\subsection{The set up:}
\begin{figure} 
\centering \includegraphics[height = 5.8cm, width = 7.8cm]{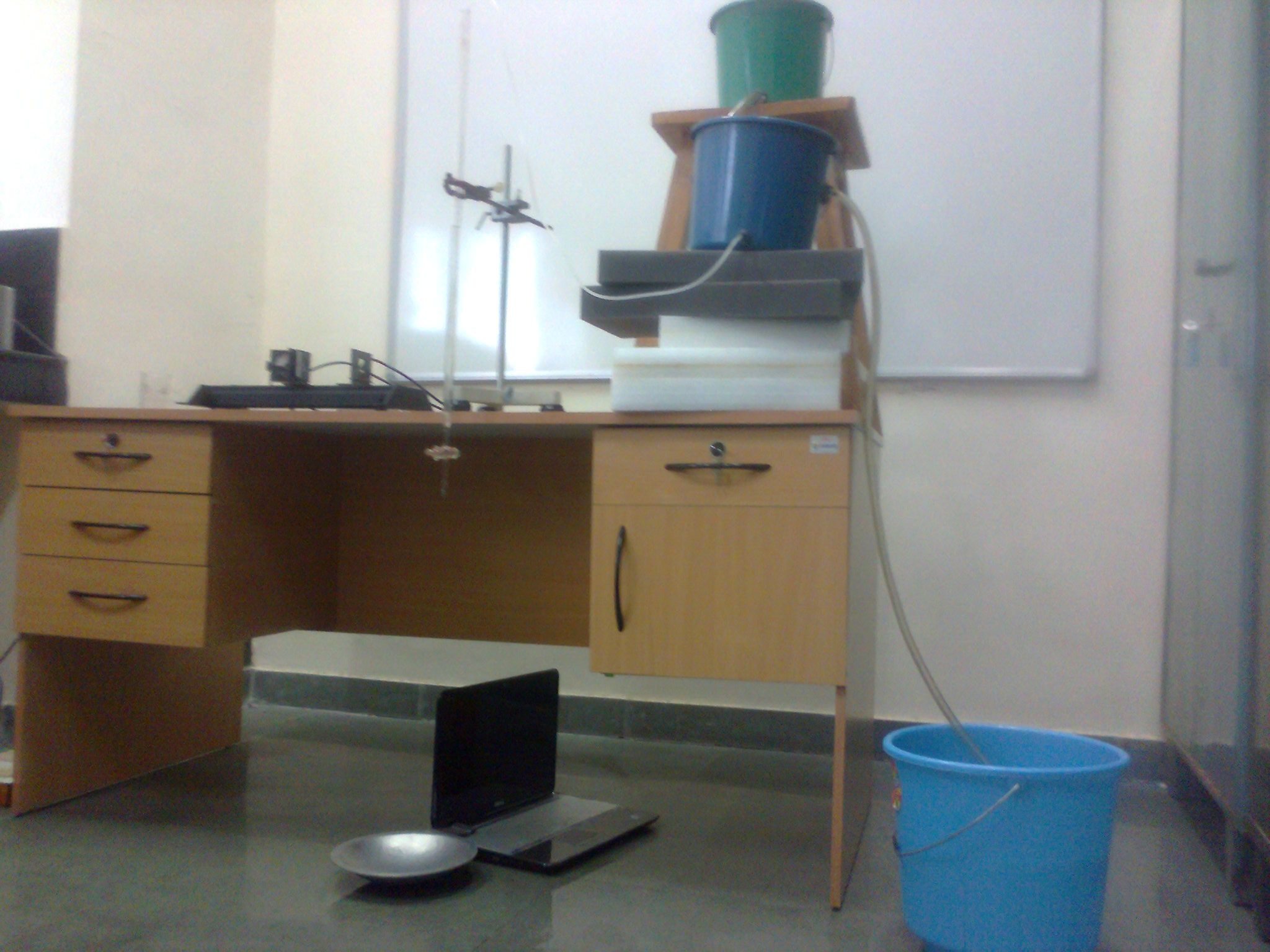}
\caption{The experimental setup showing the reservoir system and the system for creating and transmitting the drop sound.} 
\label{fig_setup}
\end{figure}

\begin{figure} 
\centering \includegraphics[height = 5.8cm, width = 7.8cm]{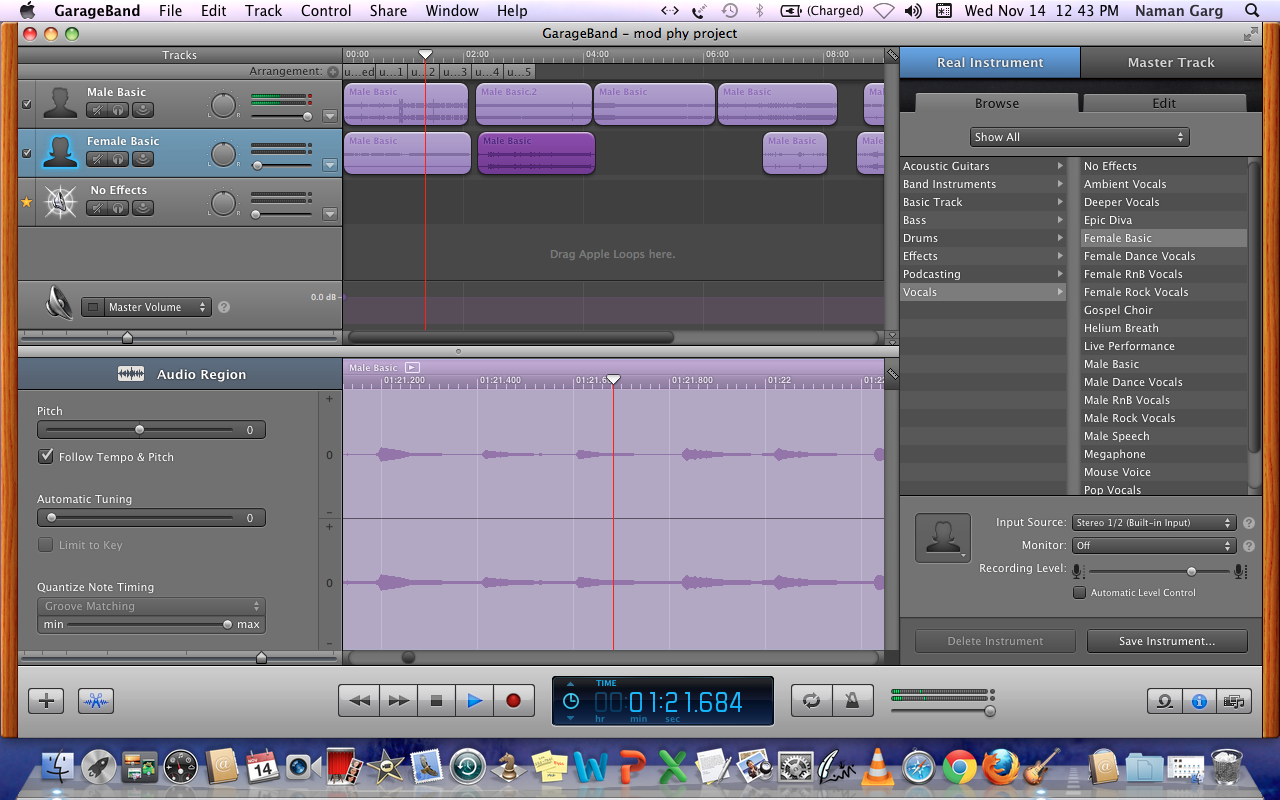}
\caption{The screen shot showing the recording on the audio analysis software interface.The drops can be clearly seen on the track shown.} 
\label{fig_audio_shot}
\end{figure}

\noi
To perform a recording of the drop falls we need to first set up a water reservoir and feeder system which maintains a constant height in the burette, which we use to create the water drops. The burette key is used to control the water flow rate $v$ which we use to label a particular flow regime. As is evident from fig.[\ref{fig_setup}] the hydrostatic level is maintained by letting water flow out from the second bucket above a certain height while water is fed to it constantly from the top bucket. The drops created impinge on a metal bowl which creates a noise which we then transmit using a metal strip as a wave guide to the microphone of our computer which records the noise (fig.[\ref{fig_audio_shot}]).
\subsection{Data Acquisition}
The noise from the drops are recorded on a track and we produce six flow regimes from which we have analysed the three flows which gave us the three characteristic regimes namely perfectly correlated, slight variation from perfect correlation and a distinct variation. The noise itself has a distribution (in time) on the track and to reduce ambiguities we have taken the beginning of the waveform as the time when the drop starts to impinge on the bowl. The times were then recorded on a spreadsheet. We took an average of one minute per run acquiring around 200-300 (for the regimes which we analyse we had 186, 344, 456) data points per run.
\section{Data analysis}
\label{sec_data_ana}
\begin{figure} 
\centering \includegraphics[height = 5.8cm, width = 7.8cm]{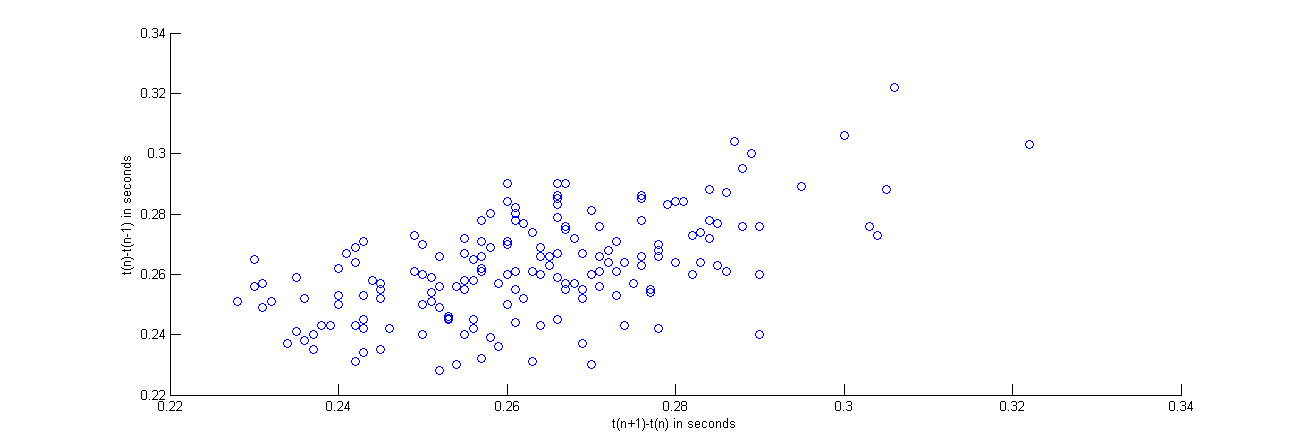}
\caption{Case 1: The scatter plot of the time differences between successive drops for a flow rate of $3.108$  drops per second$.158$ ml per second  showing almost perfect correlation and located mainly on the $y=x$ line.} 
\label{fig_case1}
\end{figure}

\begin{figure} 
\centering \includegraphics[height = 5.8cm, width = 7.8cm]{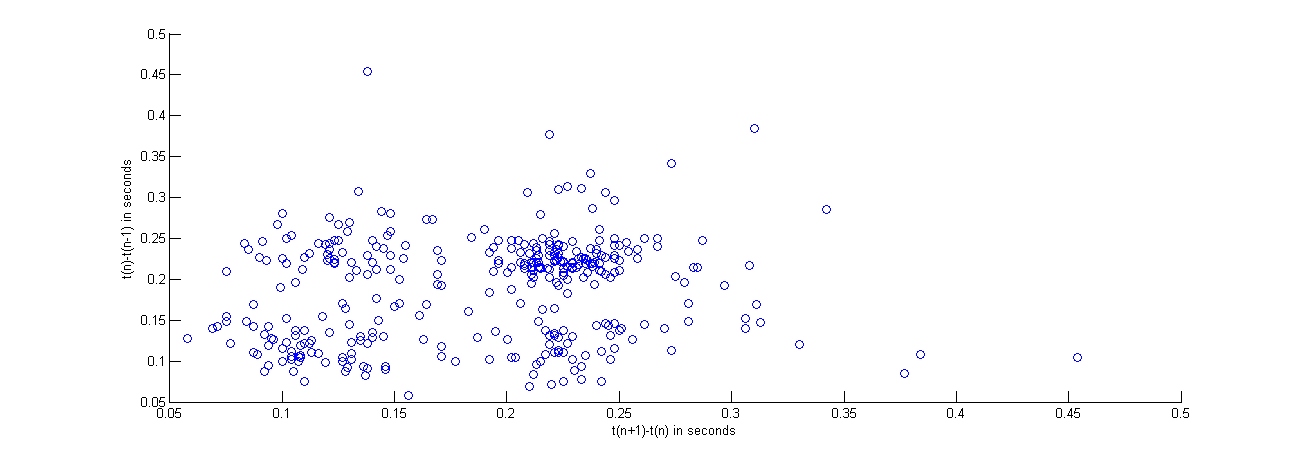}
\caption{Case 2: The scatter plot of the time differences between successive drops for a flow rate of $5.341$ drops per second$.223$ ml per second showing some amount of correlation primarily grouped into 4 different patches.}
\label{fig_case2} 
\end{figure}

\begin{figure} 
\centering \includegraphics[height = 5.8cm, width = 7.8cm]{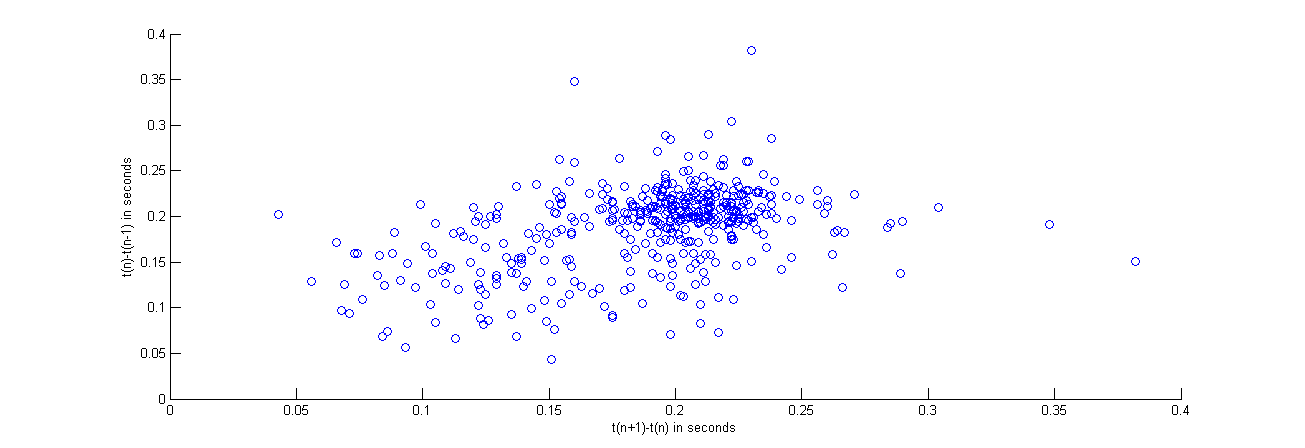}
\caption{Case 3: The scatter plot of the time differences between successive drops for a flow rate of $5.587$ drops per second $.296$ ml per second showing a lessening degree of correlation compared to case 2 between the drop times, this has the greatest $\beta$}
\label{fig_case3}
 \end{figure}

\noi
To initially characterize the flows we do a scatter plot of $ t_{n+1} - t_{n}$(y) with $ t_{n} - t_{n-1}$(x),(Figs.\ref{fig_case1},\ref{fig_case2},\ref{fig_case3}) where t is the time at which the drop impinges. For completely periodic and predictable flows we get a very dense scattering of points around the $y = x$ (fig.\ref{fig_case1}) while for the chaotic one we actually get separated clusters of such points. This is where our project gets musical, to find out regimes which are chaotic we turn the drop time data into frequencies by taking the time differences between successive drops and using $f = 1/(\delta t)$, and scaling it up to audible frequency range. We then adjust these frequencies to pentatonic major note scale in C Sharp, which enables us to relate the frequencies with our intuitive understanding of music. We do so by taking the note which is nearest to the frequency we record. We then assign a time for the note and play the `music' created. We use MATLAB to create and play these tones.
\subsection{Musical Analysis}
\noi
In the next part we listened to the tracks thus produced and chose three tracks corresponding to three flow rates which characterize the totally predictable, chaotic and random noise regimes, we then do a power spectral analysis of the data obtained to get the $\beta$ parameter after a linear fit to the data points. To obtain the power spectral density we calculate the integrals eqn.\ref{eq.h} and eqn.\ref{eq.s_v}, namely the auto correlation function $h(t)$ and its Fourier transform, taking the discrete voltage signal we have created using step functions where the voltage signal is taken to be one when the drop impinges and zero elsewhere. The integration was coded in MATLAB and we include the code snippet in the appendix (\ref{sec_Ap2}). We plot $log(S_{v}(f))$ against $log f$ to obtain the distribution needed to calculate $\beta$ which we obtain from a linear fit.
\begin{figure} 
\centering \includegraphics[height = 5.8cm, width = 7.8cm]{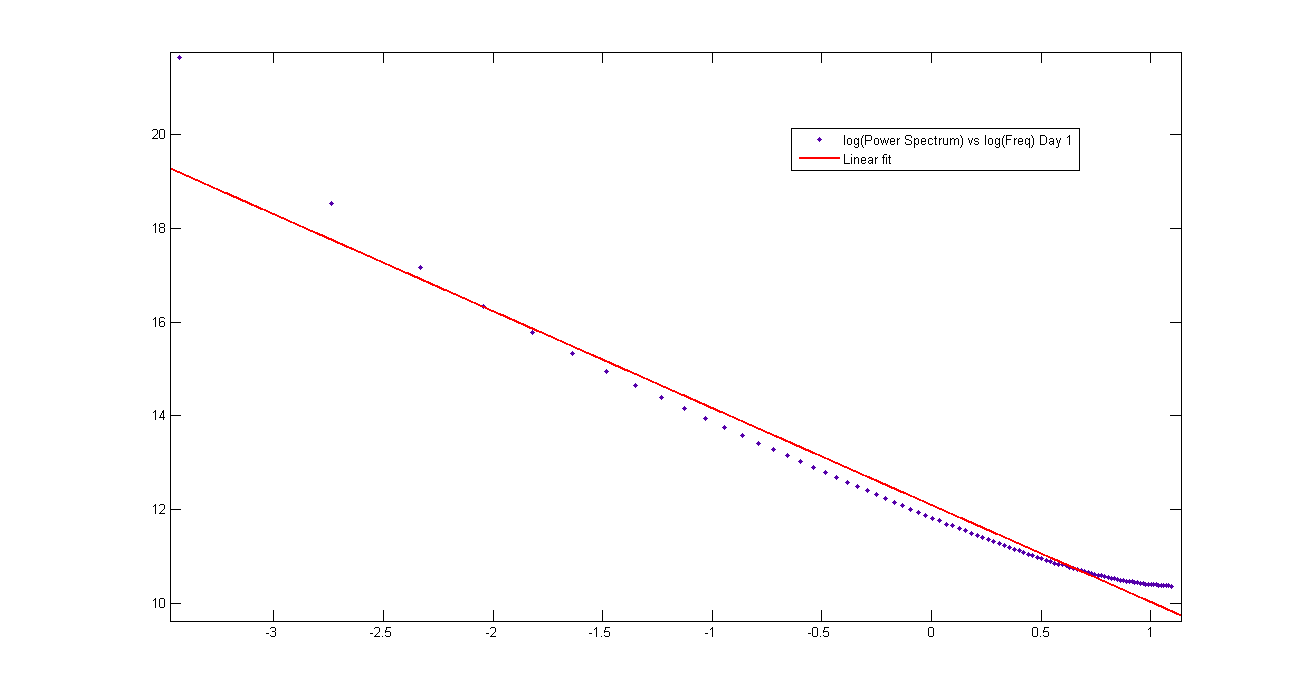}
\caption{Case 1: The $S_{V}(f)$ for case 1 showing a linear fit on a logarithmic scale.} 
\label{fig_day1}
\end{figure}

\begin{figure} 
\centering \includegraphics[height = 5.8cm, width = 7.8cm]{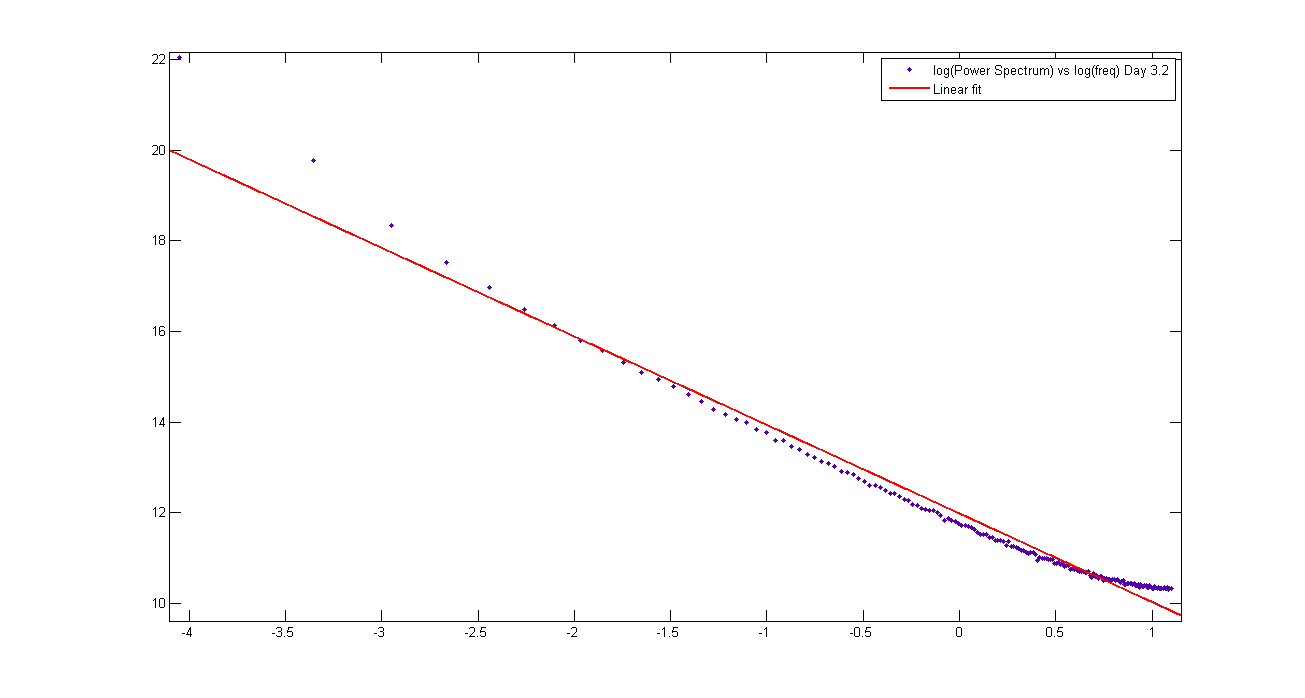}
\caption{Case 2: The $S_{V}(f)$ for case 2 showing a linear fit on a logarithmic scale.}
\label{fig_butterfly} 
\end{figure}
\begin{figure} 
\centering \includegraphics[height = 5.8cm, width = 7.8cm]{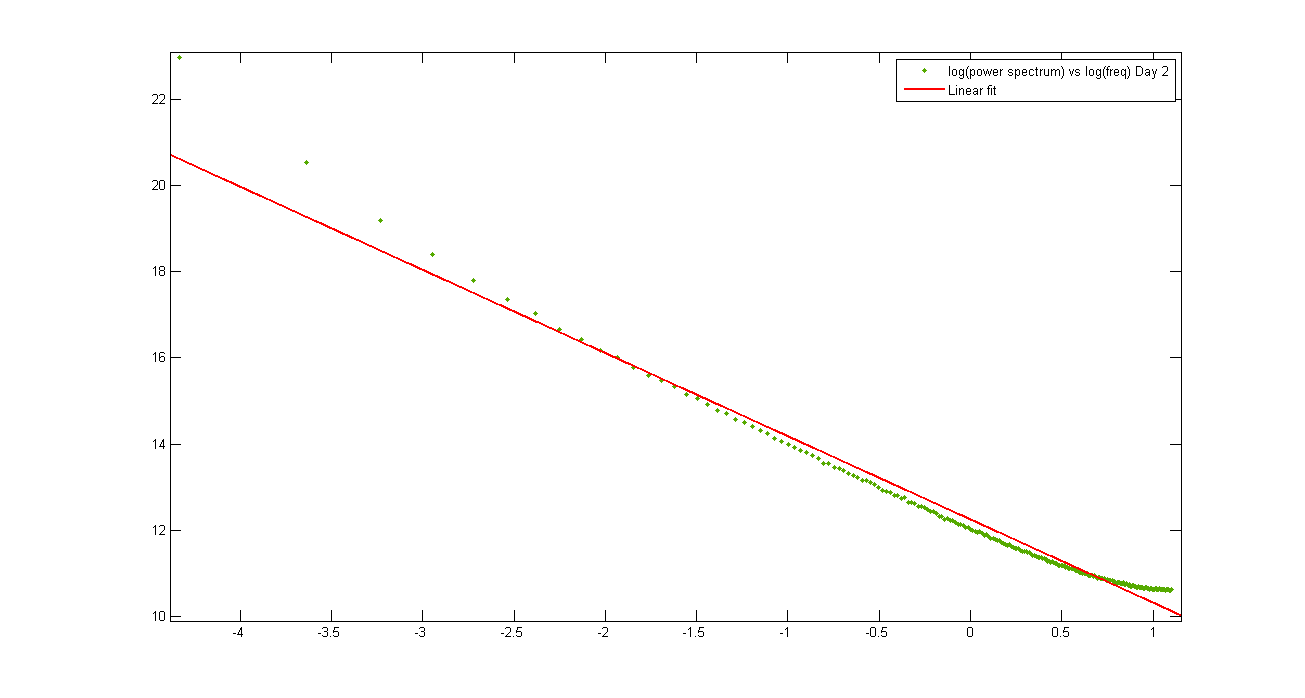}
\caption{Case 3: The $S_{V}(f)$ for case 3 showing a linear fit on a logarithmic scale.}
\label{fig_day2} 
\end{figure}
\begin{figure} 
\centering \includegraphics[height = 5.8cm, width = 7.8cm]{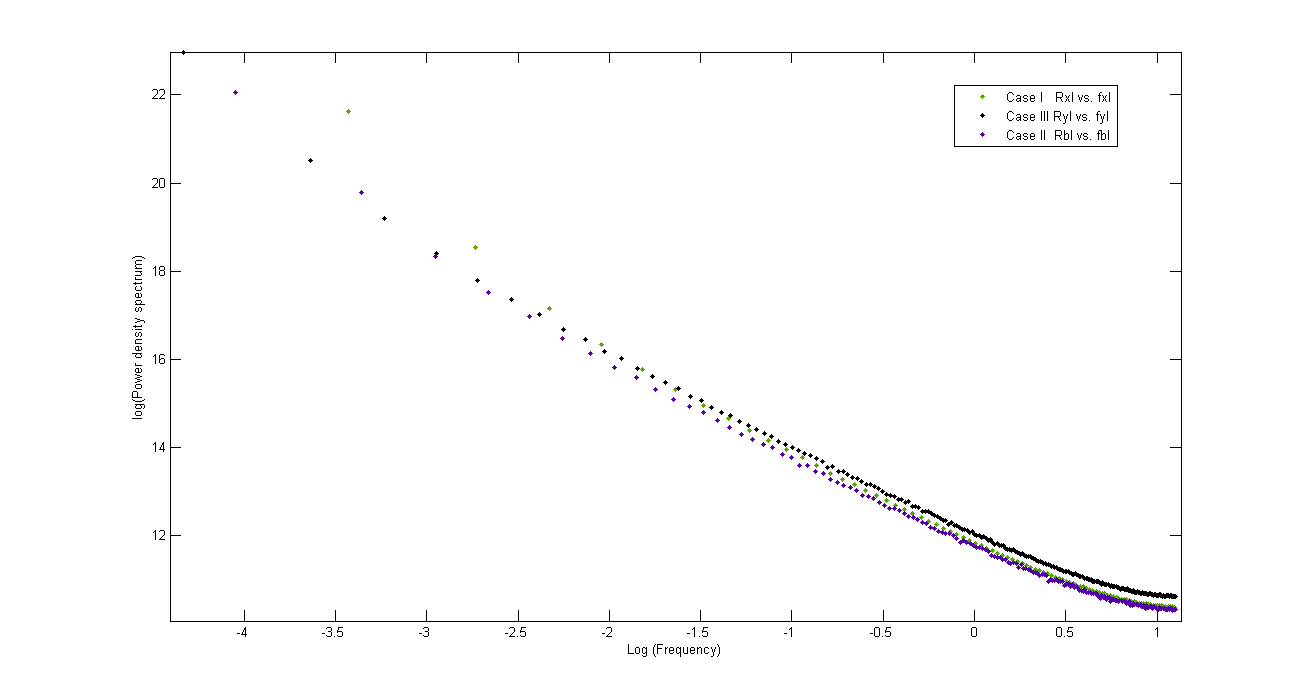}
\caption{ The $S_{V}(f)$ comparison for all the three cases. The colour scheme is Green: Case1, Purple: Case2, Black: Case3.}
\label{fig_allthree} 
\end{figure}
\subsection{Linear Fit Analysis:}
\noi
The fit has been done according to the polynomial $f(x)= p_{1}x + p_{2}$. We are reporting the results within a $95$ percent confidence bound.
\subsubsection*{Case 1, fig.\ref{fig_day1}, $3.108$  drops per second,$0.158$ ml per second):}
\noi
$p_{1} = -2.065 (-2.148,-1.981)$\\
$R^{2}=.9638$
\subsubsection*{Case 2, fig.\ref{fig_butterfly}, $5.341$ drops per second,$0.223$ ml per second):}
\noi
$p_{1} = -1.954 (-2.001,-1.908)$\\
$R^{2}=.9759$
\subsubsection*{Case 3, fig.\ref{fig_day2}, $5.587$ drops per second,$0.296$ ml per second):}
\noi
$p_{1} = -1.928 (-1.966,-1.89)$\\
$R^{2}=.9779$
\section {Error:}
\label{sec_error}
\noi
At the outset we admit that the set up we used is not the ideal one and we had initially wanted to use a laser beam interception technique to register the drops as voltage signals which would have yielded far more accurate data, however we could not do so because our attempts at setting up an oscilloscope we had in the laboratory failed because it could not store data over a sufficient period of time and also did not transmit real time voltage data to a spreadsheet software. \\
Thus, we altered our approach and decided to use sound. To create a recordable noise we had to maintain a large height between the faucet and the metal bowl and in doing so the drops were affected by air flow. We also faced the problem of background noise which we removed using noise gates and noise filters in the recording software, to this end the metallic nature of the bowl helped in producing a tone with a high enough frequency and amplitude which could be distinguished from background noise by the software. We created a metallic guide from the bowl to the microphone of our computer to further enhance the signal. These sources of error are not mathematically quantifiable and the only error we quantify is the statistical error we obtain by fitting the frequency and spectral power plots to a linear fit.
\subsection{Results:}
\label{sec_results}
\noi
$\beta_{1} = 2.065 (2.148,1.981)$\\
$\beta_{2} = 1.954 (2.001,1.908)$\\
$\beta_{3} = 1.928 (1.966,1.89)$\\
\section{Conclusion}
\label{sec_conclusion}
\noi
From the linear fits for the three cases we analyse, we find that the data set which had the least amount of correlation between the points has the smallest $\beta$ which we take to be the parameter relating the data set to its musical quality. The data set which is most correlated does indeed produce a $\beta$ factor of two that is $S_{V}(f)=\frac{1}{f^{2}}$ indicating according to the theory outlined that it falls within the regime of Brownian noise.\\
 To make the music more familiar we added overtones to the base frequency, obtained from the $\delta t$, $f'$ by adding frequencies $2f'$ and $3f'$ with amplitudes $0.5$ and $0.33$. This $f'$ is obtained by fitting the frequencies we obtain from the time differences to the nearest note in the pentatonic major scale of music.\\
 The data we obtain agrees with the power law in that a near perfect correlation does give the Brownian noise relation and a lower $\beta$ sounds more musical if not a lot. More detailed studies of the power law on musical compositions have revealed $\beta$ ranges of $0.8$ to $1.4$.
\section{Acknowledgement}
\label{sec_Ack}
\noi
We would like to thank our supervisor Dr A. Gupta for several useful discussions and Mr Prashar of the Modern Physics Laboratory, IIT Kanpur for helping procure equipment for the experiment and sorting out various issues related to their proper functioning. We also thank our classmate Venkat Kapil who suggested the pentatonic scale for creating the palatable music.

\section{Appendix $1.1$ (Computer codes for music):}
\label{sec_Ap1}
\noi
The code below first converts the array of frequencies corresponding to the time intervals between drops scaled up to audible frequencies b[$344$] to bp[$344$] which is a new array with frequencies on the pentatonic major scale in C sharp closest to the original frequencies. Then it plays it using the function soundsc where we create a tone by adding the harmonics of the given frequency. For example for Case II-\\
Butterfly pentatonic\\
\texttt{i=1
j=1\\
 for$ j = 1:344$\\
    diff$=100000$\\
 for i = $1:20$\\
 diffn = abs(b(j)-notep(i))\\
if diffn$>=$diff\\
 bp(j)=notep(i-$1$)\\
bpi(j)=i-$1$\\
break\\
else\\
  diff= diffn\\
  end\\
  i=i+1\\
 end\\
 j=j+1\\
end\\
i=1 \\
for i = 1:344\\
     song =$\sin(2*\pi*bp(i)*(0:0.000125:0.15)) + 1/3*\sin(2*2*\pi*bp(i)*(0:0.000125:0.15)) + 1/5*\sin(2*3*\pi*bp(i)*(0:0.000125:0.15))$ \\
     soundsc(song)\\
     i=i+$1$\\
end}
\section{Appendix $1.2$ (Computer codes for $S_{v}(f)$):}
\label{sec_Ap2}
\noi
The code below takes the array of time event when the drops are recorded (bt), Fourier transforms it and the power spectral density is the modulus square of the Fourier transform appropriately scaled.\\
\texttt{x=bt;\\
N=$344$;\\
fs=$6$;\\
fh=$3$;\\
xf=fft(x);\\
Rb=xf.*conj(xf)/N;\\
k=fh*N/fs;\\
fb=fs/N*(1:N/2);}\\

\end{document}